\documentstyle[prl,aps,epsf,preprint]{revtex}

\begin{document}
\tighten

\title{GLUON CONDENSATE FROM LATTICE QCD\thanks
{This work is supported in part by funds provided by the U.S. Department
of Energy (D.O.E.) under cooperative agreement \#DF-FC02-94ER40818.}}

\author{Xiangdong Ji }

\address{Center for Theoretical Physics \\
Laboratory for Nuclear Science \\
and Department of Physics \\
Massachusetts Institute of Technology \\
Cambridge, Massachusetts 02139 \\
{~}}

\date{MIT-CTP-2439 \hskip 1in  hep-ph/9506413 \hskip 1in June 1995}

\maketitle

\begin{abstract}

After making some critical comments about the traditional method of
extracting the gluon condensate from lattice QCD data, I present an
alternative analysis.  The result is more than a factor of five larger
than the phenomenological value.  Two closely related subjects, the
effects of the infrared renormalons on the extraction and the Lepage and
Mackenzie improvement on a lattice perturbation series, are also
discussed.

\end{abstract}

\pacs{xxxxxx}

\narrowtext
One of the popular approaches to study
Quantum Chromodynamics (QCD) in the strong coupling
region is the QCD sum rule method initiated by
Shifman, Vainshtein, and Zakharov (SVZ) more than
fifteen years ago \cite{SVZ}. The success of the approach
has been demonstrated by many examples ranging from
hadron masses, light-front wavefunctions to
decay widths, form factors, etc. Central to the
QCD sum rules is the concept of vacuum condensates
which are expectation values of composite operators
in the QCD vacuum. Among those, the most legend
is perhaps the gluon condensate defined through
the operator $F^2=F^{\alpha\beta}F_{\alpha\beta}$,
where $F^{\alpha\beta}$ is the strength of color
gauge fields.

To understand the sum rule phenomenology at
a more fundamental level,
one has to calculate the vacuum condensates
directly from the QCD lagrangian. Calculations using
lattice Monte Carlo started in several groups
shortly after the publication of SVZ's
paper\cite{BAN}. Before I comment on
these calculations, it is important to point out
that phenomenological condensates
from fitting experimental data are in principle different
from theoretical condensates that are
calculated as matrix elements in the QCD vacuum.
The former are extracted with
the Wilson coefficients computed to
a few loops, and thus may
contain large uncalculated
multi-loop contributions from the coefficient
functions and higher-power corrections.

The traditional approach of calculating the gluon
condensate (in the quenched approximation)
goes like this \cite{BAN}.
Consider an elementary
plaquette on the lattice. Calculate using Monte Carlo
the trace of the plaquette as a function
of the lattice spacing $a$, or the lattice coupling constant
$\beta=2N_c/g^2_0$. For small enough $a$, one
has,
\begin{equation}
      1 - {1\over N_c} {\rm Tr}P
 =  \sum_{n=1} {c_n \over \beta^n} + {\pi^2\over 12 N_c}
    a^4(\beta) \langle {\alpha_s\over \pi} F^2\rangle + {\cal O}(a^6
) \ .
\end{equation}
The various terms on the right-hand side have different
characteristic $\beta$ (or $a$) dependences.
The leading term is a series logarithmic
in $a$ (or power-like in $1/\beta$). The condensate
term is quartic in $a$ (or exponential
in $\beta$). By calculating both the left-hand side
and the perturbation series for a wide range of $\beta$,
the condensate can be extracted by fitting
the expected $\beta$ dependence.

The approach shall work in principle. But it
is difficult to implement in practice. Let me list
a few of the practical problems with the approach:
\begin{itemize}
\item{First, a big lattice is required to compute
the plaquette average for small $a$.
Physically the gluon condensate comes from the effects of
long wavelength gluons (somewhere from 0.5 to
1 fm). The combination of a small lattice spacing and
a reasonable physical size requires a big lattice.
In ref. \cite{CAM}, it was determined that
the asymptotic region, where the data
behave like a sum of powers plus an exponential
in $\beta$, starts from
$\beta= 6.58$. Using a two-loop relation between
$\beta$ and $a$, $a=(\beta/6b_0)^{b_1/2b_0^2}/\Lambda_L
\exp(-\beta/12b_0)$ with $\Lambda_L=4.4$ MeV,
$b_0=11/16\pi^2$ and $b_1=102/256\pi^2$,
one finds that the coupling corresponds to
a lattice spacing 0.055
fm. Thus a lattice with 8 points in spatial
directions spans a physical dimension of
0.44 fm. For $\beta=7$, the lattice size further
reduces to 0.27 fm. Both lattices seem to be too small
to measure non-perturbative physics.}
\item{Second, relative to the leading term,
the condensate term contributes little
to Eq. (1) as $a$ decreases.
At $a=0.055$ fm, the ratio between the leading and the
condensate terms is more than $10^3$. Thus to determine the
condensate with a reasonable error at small $a$, one has to
compute the perturbation series to large orders.}
\item{Finally, the perturbation series in Eq. (1) actually
divergences due to the infrared (IR) renormalons, that is,
the coefficients of the series increase like $n!$
and with a fixed sign\cite{TH}. The presence of the renormalons
complicates the extraction of the condensate
in two ways. 1) The
perturbation series cannot yield better accuracy
beyond a certain order. 2) The difference
between power and exponential behaviors in $\beta$, which
has been the basis for fitting the condensate,
disappears to a certain degree.}
\end{itemize}

The physical significance of the gluon condensate in light of the
IR renormalons has been discussed in the literature
for a long time\cite{NOV,DAV,ULS}. The upshot of these
discussions is that the gluon condensate is a
procedure-dependent concept. [In this sense, the
status of a theoretical condensate is not much different from a
phenomenological one.] When a condensate
is calculated in a particular scheme designed to
regularize the infrared renormalons, the
coefficient functions in an operator product expansion
must be calculated accordingly.
A consistent method to regularize the renormalons both
in the coefficient functions and condensates,
generalizing the one-loop discussion in ref. \cite{ULS},
has recently been studied by this author \cite{JI}.

In light of the above observations,
I consider an alternative strategy to extract
the gluon condensate from Eq. (1).
Assuming the plaquette perturbation
series is an asymptotic one,
one would expect that the magnitude of its terms
initially decreases, then reaches a minimum,
and finally increases without bound.
It is generally believed that the minimal term
occurs at order $n \sim \beta$, with a magnitude
$\sim a^4$.  Thus the minimum uncertainty in the
perturbation series induces an uncertainty in the gluon
condensate, which can only be eliminated through
a regularization of the series.
However, if the uncertainty is small compared with the
condensate itself, as Novikov et al. argued \cite{NOV},
it still makes a good deal of physical sense to extract
a gluon condensate independent of the regularization
scheme. To illustrate this more clearly, I schematically
show in Fig. 1 the order at which the minimum term occurs
as a function of $\beta$ (the solid line labelled
by $M$). I also show the order at which the size of the
perturbative term is roughly that of the gluon condensate
(with the solid line labelled by $C$).
If the uncertainty on the condensate induced by IR
renormalons is small, the $C$ line would be significantly
below the $M$ line.

Now suppose the perturbation series has been calculated to
some fixed order $n_0$, shown by the dashed line
in Fig. 1, and call its intersection with curve $C$,
$\beta_{\rm max}$. Clearly, one cannot
extract the gluon condensate accurately
if $\beta>\beta_{\rm max}$. On the other hand,
for a small $\beta$, the corrections
from the lattice discretization and the higher-dimensional
condensates in the expansion become
important. Thus there exists a lower limit on $\beta$,
$\beta_{\rm min}$, below which the extraction
becomes unreliable. The existence of a window in $\beta$
($\beta_{\rm min}
<\beta_{\rm max}$) depends on $n_0$.
So long as a window exists, one shall be able to
make an approximate extraction.
The error on the condensate depends on
$n_0$. When $n_0$ is greater than
$n(\beta_{\rm min})$ on the $M$
curve, the error is limited by the renormalon singularity.

\goodbreak
\vskip12pt
\nobreak
\centerline{\epsffile{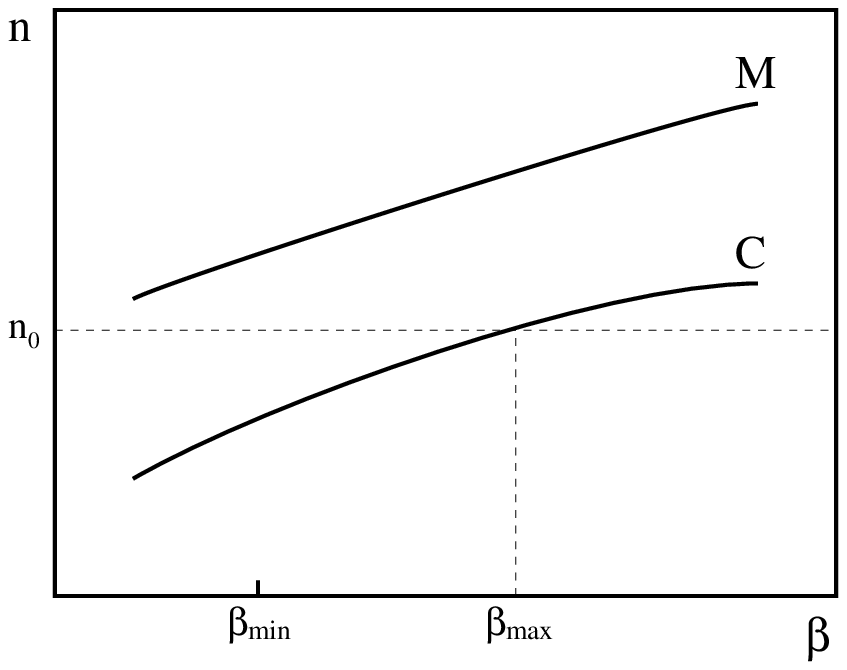}}
\nobreak
\centerline{Fig.~1.~~  See the text
for the explanation of various curves.}
\nobreak
\vskip12pt
\goodbreak

The perturbation series for the
elementary plaquette has initially been calculated to
three loops by Alles, Campostrini, Feo, and Panagopoulos \cite{ALL}
and recently to eight loops
by Di Reno, Onofri, and Marchesini using a numerical
method \cite{DR}. With $n_0 = 8$,
$\beta_{\rm max }$ is somewhere around 6.5 to 7.
Experiences with lattice calculations
indicate a $\beta_{\rm min}$ around 5.7.
In the following analysis,
I choose three lattice couplings:
$\beta=5.7$, 6.0, and $6.2$, corresponding to
the two-loop lattice spacings $a = 0.15$, 0.105, 0.084 fm, respectively.

For the average of the elementary plaquette, I use
the following result from the lattice Monte Carlo \cite{CAM,BMR,LP},
\begin{eqnarray}
  1 - {1\over N_c} {\rm Tr}P &&= 0.4509 ~~ (\beta = 5.7) \ ,\nonumber \\
         &&= 0.4058 \ \  (\beta = 6.0) \ ,  \nonumber  \\
         &&= 0.3861  \ \  (\beta = 6.2) \ .
\end{eqnarray}
Since the numerical errors are small compared with
the uncertainty in perturbation series, I have ignored them. The perturbation
series from Ref. \cite{DR} is,
\begin{equation}
     \sum_n {c_n\over \beta^n}= {2\over \beta}  +  {1.218\over \beta^2}
    +  {2.960\over \beta^3}
    +  {9.28\over \beta^4}  +  {34\over \beta^5} +  {135\over \beta^6}
    +  {563\over \beta^7} +  {2488\over \beta^8} + ...
\label{pert}
\end{equation}
where I have also ignored the small Monte Carlo errors.
For $\beta$'s under consideration,
the last term calculated is not yet the minimal term in the
series. I assume for the moment the error in truncating the series
is determined by the last term,
\begin{eqnarray}
      \sum_n {c_n\over \beta^n} && =   0.4278 \pm 0.0022 \ \
         (\beta=5.7) \ , \nonumber \\
               &&= 0.3988 \pm 0.0015  \ \
     (\beta=6.0) \ , \nonumber \\
               && = 0.3818 \pm 0.0011 \ \  (\beta=6.2) \ .
\end{eqnarray}
Then the differences between the Monte Carlo data and
the perturbation series are, $0.0222 \pm 0.0022$,
$0.0075 \pm 0.0015$, $0.0045 \pm 0.0011$ for $\beta=5.7$,
$6.0$, $6.2$, respectively. In producing the small differences,
the knowledge of four, five, and six-loop
terms in the series has been essential.
Notice that the relative error
is larger for larger $\beta$. This reflects
the observation made early---to extract the condensate at
larger $\beta$, one needs higher
precision for the series.

The error estimate in Eq. (4) is probably optimistic.
One indication is
that terms in the perturbation series have the same sign
and decrease slowly with increasing $n$. To improve
the extraction, one may estimate
higher-order terms in the series.
In the following, I shall consider two such estimates:
the Pade approximation and the leading IR renormalon
approximation.

In the standard Pade approach, both [3,4] and [4,3]
approximations produce ([m,n] is the standard notation
for the Pade approximation with a $m$-th order polynomial in the
numerator and a $n$-th order polynomial in the denominator),
\begin{eqnarray}
      \sum_n^\infty {c_n\over \beta^n} && =   0.4322  \ \
         (\beta=5.7) \ , \nonumber \\
               &&= 0.4011   \ \
     (\beta=6.0) \ , \nonumber \\
               && = 0.3833 \ \  (\beta=6.2) \ .
\end{eqnarray}
The differences between the Monte Carlo data and
the Pade approximation
are, $0.0187$,
$0.0047$, $0.0028$ for $\beta=5.7$,
$6.0$, $6.2$, respectively. The numbers are about two standard
deviations away from the eight-loop result, indicating that
that the eighth-order term may not be in the asymptotic
region.

Using the leading IR renormalon series constructed
in Ref. \cite{DR}, I find the minimal term for the plaquette
series occurs somewhere around the 25th to 30th order for
the present $\beta$'s. The first few terms beyond
the eighth order go like this,
\begin{eqnarray}
      \sum_n^\infty {c_n\over \beta^n}
    = &&  ...
   + {1.1\times 10^4 \over \beta^9}
   + {4.8\times 10^4 \over \beta^{10}}
   + {2.1\times 10^5 \over \beta^{11}}
   + {9.5\times 10^5 \over \beta^{12}}
   + {4.2\times 10^6 \over \beta^{13}}  \nonumber \\
   && + {1.9\times 10^7 \over \beta^{14}}
   + {8.9\times 10^7 \over \beta^{15}}
   + {4.1\times 10^8 \over \beta^{16}}
   + {1.9\times 10^9 \over \beta^{17}}
   + {9.4\times 10^9 \over \beta^{18}} + ...
\end{eqnarray}
Summing
the series in the renormalon approximation to the minimal term
yields,
\begin{eqnarray}
      \sum_n {c_n\over \beta^n} && =   0.4362  \ \
         (\beta=5.7) \ ,  \nonumber \\
               &&= 0.4029   \ \
     (\beta=6.0) \ , \nonumber \\
               && = 0.3848  \ \  (\beta=6.2) \ .
\end{eqnarray}
The differences between the Monte Carlo data and
the perturbation series in the leading-renormalon approximation
are, $0.0147 $,
$0.0029$, $0.0013$ for $\beta=5.7,
6.0, 6.2$, respectively. The numbers are about three standard
deviation away from the sum to eight loops.

As a final result, I take the average of the
two large-order estimates as
the central value and take their difference
as the error estimate. Thus
the condensation contributions to the
plaquette average are $0.0167 \pm 0.0040$,
$0.0038 \pm 0.0018$, and $0.0021 \pm 0.0015$ for $\beta=5.7,
6.0, 6.2$, respectively.
These numbers can be converted
to the standard form of the gluon condensate in unit of GeV$^4$,
\begin{eqnarray}
    \langle{\alpha_s\over \pi}F^2\rangle
   && = 0.18 \pm 0.04 \ \ (\beta=5.7) \ , \nonumber  \\
   && = 0.17 \pm 0.08 \ \ (\beta=6.0) \ , \nonumber \\
   && = 0.23 \pm 0.17 \ \ (\beta=6.2) \ .
\end{eqnarray}
Again, the large error for $\beta=6.2$ indicates that
the perturbation series must be calculated to higher orders
in order to extract the condensate at large $\beta$.

The numbers above are consistent with
that extracted in Ref. \cite{CAM}. They are
at least a factor of five larger than the phenomenological
determination (See Ref. \cite{DOS} for a discussion
about the phenomenological condensate).
Several factors may contribute to this discrepancy: 1)
higher-orders in the plaquette series are different
from what we expected, 2) quenched
approximation, 3) underestimation of the
lattice spacing, 4) the phenomenological
condensate is contaminated by higher-order
Wilson's coefficients.  Clearly, it is
interesting and also important to resolve
the discrepancy in this high-precision confrontation
between lattice Monte Carlo, QCD perturbation theory, and
hadron phenomenology.

The perturbation series
for the elementary plaquette (Eq. (\ref{pert}))
is the first that has been calculated
up to more than two-loops in lattice QCD. Consequently,
it is interesting to explore its convergence
properties. In general, because of the large tadpole
contributions, lattice perturbation series
converges very slowly. The plaquette series
does show such slow convergence due to an abnormal fast increase
in the coefficients, as observed
by Di Renzo et al. \cite{DR}. If a perturbation
series has been calculated
to sufficient large orders, the slow convergence is not a problem.
In
fact, as I will argue late that the lattice perturbation series
has the virtue of producing small renormalon uncertainties.

Lepage and Mackenzie \cite{LP} have recently
pointed out that the perturbative
expansion with a more "physical" coupling
accelerates the convergence of a lattice series.
Clearly, if a series is limited to the first
few terms, the acceleration
improves its predictive power considerably.
The plaquette series provides an excellent example
for studying the effects of acceleration.
To show this, let me introduce
\begin{equation}
     F = {\beta\over 2}(1-{1\over N_c}{\rm Tr} P)-1 \ .
\end{equation}
The series for $F$ is,
\begin{eqnarray}
    F =&& 1.2755~\alpha_{s0} + 6.4920~ \alpha_{s0}^2
     + 42.6276~ \alpha_{s0}^3 + 327.1~\alpha_{s0}^4 \nonumber \\
     && + 2720.1~ \alpha_{s0}^5 + 23758.8~ \alpha_{s0}^6
     +219899~\alpha_{s0}^7 + ... \ ,
\end{eqnarray}
where $\alpha_{s0}$ is the bare lattice coupling.
If truncating to
the leading term, the result for $F$, 0.1015, is
only half of the full series, 0.2087, at $\beta=6.0$.
The bad approximation by the leading term is related to
the large coefficient 6.4920 in the second term and,
in general, the fast increase of the perturbative
coefficients. However, If changing to a
renormalized coupling which coincides
with the one-loop $\overline{\rm MS}$ coupling at scale
$\mu = \pi/ a$,
\begin{equation}
     \alpha_{s0} = \alpha_{s}/( 1+3.88~\alpha_{s})   \ ,
\end{equation}
$F$ becomes,
\begin{equation}
    F = 1.2755~\alpha_{s} + 1.543~ \alpha_{s}^2
     + 11.44~ \alpha_{s}^3 + 49.7~ \alpha_{s}^4
     + 265.8~ \alpha_{s}^5 + 1566~ \alpha_{s}^6
     + 9616 ~\alpha_{s}^7 + ... \ .
\label{li}
\end{equation}
Clearly, the convergence has been improved considerably
due to the small expansion coefficients. The leading term is now 0.1584,
a number much closer to the full result. This phenomenon has already
been studied in Ref. \cite{DR}.

Two comments can be made about the acceleration of the
lattice perturbation series, both of which are
related to the question of precision, important, as we
have seen, in extracting
the gluon condensate. First, in the expansion with a new
coupling, the coefficients of the lower-order terms
in general do not increase at a monotonic rate, i.e., they
become less regular. Regularity
is a necessary (not sufficient, of course)
diagnosis for the series in the asymptotic
region, where one can estimate the error of truncation
using the last term included.
To produce a series with regular
lower-order terms, one must use some "natural"
couplings with well-defined relation to
the lattice coupling up to high orders.
One possibility is to use couplings
associated with some "physical" schemes
like $\overline{\rm MS}$. However, the relation
between the lattice and $\overline{\rm MS}$ couplings
is difficult to calculate. Only recently, the relation
has been computed to two-loops by
Luscher and Weisz \cite{LW},
\begin{equation}
   \alpha_{s0} = \alpha_s -3.88~\alpha_s^2 +
              7.744~\alpha_s^3 + .... \ ,
\end{equation}
where $\alpha_s$ is again defined at scale $\mu=\pi/a$.
Using a rationalized version of the above relation, $\alpha_{s0}
= \alpha_s/(1+3.88\alpha_s +7.310\alpha_s^2)$,
I have,
\begin{equation}
    F = 1.2755~\alpha_{s} + 1.543~ \alpha_{s}^2
     + 2.128~\alpha_{s}^3 + 29.20~ \alpha_{s}^4
     + 169.0~ \alpha_{s}^5 + 880.5~ \alpha_{s}^6
     + 9835~ \alpha_{s}^7 + ... \ .
\end{equation}
The coefficient of $\alpha_s^3$ term is now
reduced significantly, compared with that in Eq. (\ref{li}). However,
the irregularity has been shifted to higher-order terms.
This is reflected by the fluctuation of
the successive terms, which makes
error-estimate difficult.

The second comment is about the intrinsic error
of the series due to IR renormalons. In new expansions,
the series converges faster, however, the minimal term in
the series in general becomes larger. Take the above series as an
example, the smallest term
in the series is approximately 0.0030.
Before the improvement, however, the
smallest term in the series is less than
$10^{-4}$. This dependence of the renormalon
uncertainty on expansion schemes is not well
studied and is quite interesting. It may imply
the existence of a scheme
in which the coupling constant approaches to zero and the
power series converges infinitely slowly, but,
the renormalon ambiguity may be vanishingly small.

To summarize, I have critically examined the
traditional method of extracting the
gluon condensate from the lattice QCD data.
Using the plaquette series calculated recently
to eight loops, I have made a direct
determination of the condensate at low
$\beta$ and large lattice spacings.
The result is at least a factor of five larger than
the phenomenological condensate.
Although still higher-order terms in the perturbation
series must be computed to reduce the uncertainty,
the intrinsic error due to IR
renormalons seems to be small. The plaquette series
converges faster when expanded with
more physical couplings, however, the
estimation of errors becomes more difficult.

\acknowledgments
I thank G. Grunberg for bringing my attention to the
recent eight-loop result and G. Marchesini for
a conversation about the perturbative calculation.
I also thank A. Di Giacomo for references
on lattice determinations of the gluon condensate
and J. Negele for useful discussions.

\end{document}